\documentclass[prl,aps,twocolumn]{revtex4}

\usepackage{amssymb}
\usepackage{amsmath}
\usepackage{epsf}
\usepackage{psfrag}
\usepackage{graphicx}
\usepackage{bm}

\def\pa{{\partial}}

\def\imag{\dot \imath}
\begin{document}

\author{Ana Ach\'{u}carro}
\email[Electronic address: ]{achucar@lorentz.leidenuniv.nl}
\affiliation{Instituut-Lorentz for Theoretical Physics, Leiden,
The Netherlands}
\affiliation{Department of Theoretical Physics, The University of the Basque Country UPV-EHU, 38940 Bilbao, Spain}
\author{Roland de Putter}
\email[Electronic address: ]{rdeputter@lorentz.leidenuniv.nl}
\affiliation{Instituut-Lorentz for Theoretical Physics, Leiden,
The Netherlands}

\title{Effective non-intercommutation of local cosmic strings at high collision
speeds} \date{\today}

\begin{abstract}
We present evidence that Abrikosov-Nielsen-Olesen (ANO) strings pass
through each other for very high speeds of approach due to a double
intercommutation. In near-perpendicular collisions numerical
simulations give threshold speeds bounded above by $\sim 0.97 c$ for
type I, and by $\sim 0.90 c$ for deep type II strings.  The second
intercommutation occurs because at ultra high collision speeds, the
connecting segments formed by the first intercommutation are nearly
static and almost antiparallel, which gives them time to interact and
annihilate. A simple model explains the rough features of the
threshold velocity dependence with the incidence angle.  For deep type
II strings and large incidence angles a second effect becomes
dominant, the formation of a loop that catches up with the
interpolating segments. The loop is related to the observed vortex -
antivortex reemergence in two-dimensions. In this case the critical
value for double intercommutation can become much lower.
\end{abstract}

\maketitle

Cosmic strings were intensely studied in the eighties as a possible
explanation of the small deviations from homogeneity that are
necessary to seed structure formation in the early Universe. Since
then, this picture
has been abandoned in favor of the inflationary scenario, mainly because of
observations of the cosmic microwave background
radiation \cite{PogWymWass04}.  The recent revival of interest in
cosmic string is largely motivated by fundamental theory.  First, the
formation of cosmic strings in the early universe appears to be a
fairly generic prediction of grand unified theories of particle
physics \cite{jeann03}. Second,
some brane inflation
models from superstring theory predict the formation of cosmic
(super)string networks as well \cite{SarTye02,MajDavis02,CopMyersPolch04,DvaliVil04}.
It should be possible in the near future, for example using B modes of
the CMB polarization \cite{SelSlos06}, to detect the effects of
cosmic strings with
a sensitivity one or two orders of magnitude higher than at present,
which makes strings an excellent probe of physics at ultra-high
energies that are otherwise extremely hard to probe (even the absence
of strings can discriminate between particle physics
models).

\vspace{0.5cm}

An important property of cosmic strings is their behavior when two
string segments collide. In principle, there are four
possible outcomes (we refer the reader to \cite{VilShell94} for general
background and references): they can
(1) simply pass through each other, (2) exchange ends and
reconnect (3) form a Y-junction with a bridge between the
original strings or (4) do neither and get tangled up,
which happens if the Y-junction is kinematically forbidden
\cite{CopKibbSteer06}.
Outcomes (3) and (4) apply to non-abelian gauge theory strings, where
there is a topological obstruction that forbids the first two
outcomes, and also to $(p,q)$ strings, which are bound states of
fundamental superstrings and $D1$-branes. Another example of outcome
(3) are the zipper bound states that can be formed between
multiply-winding type I ANO vortices when they collide at low speeds
and incidence angles, due to their attractive
interaction~\cite{BettLagMatz97}.

\vspace{0.5cm}

The second outcome is usually called {\it intercommutation} (or
reconnection) and is extremely important in cosmological scenarios as
it provides a mechanism for a string network to lose energy and reach
a scaling regime  in which the energy density in strings remains a
constant fraction of the dominant form of energy density in the
Universe (matter or radiation).
Intercommutation leads to loops and small scale structure that decay
efficiently into particles and radiation.
The question of the precise effect of intercommutation on scaling (see for
example \cite{VincAntHind98,MooreShellMart02,MartMooreShell04,
JonToiTye03,DamVil05,Sak05,AvgShell06,VanchOlVil05,RingSakBouch05}) has
taken centre stage recently after the realization that for cosmic
superstrings the probability of intercommutation $p$ is very low
($p \sim 10^{-3}$ to $\sim 10^{-1}$ depending on the type of
string \cite{JackJonPolch05}). While all previous studies
agree that the network will reach a scaling solution if $p=1$,
the situation is less clear for lower intercommutation probabilities.
Since full field theory simulations of string networks on cosmological scales are computationally too demanding,
numerical studies of such networks (\cite{AlbrTurok85,Benn86}) typically use the effective Nambu-Goto action (\cite{AndBonGregStew97}), which treats strings as infinitely
thin objects. However,
this action cannot be used to describe what happens when two strings intersect and therefore the intercommutation behavior
needs to be studied using the full field theory.

\vspace{0.5cm}

\begin{figure}
  \begin{center}{
  \includegraphics*[width=5.5cm]{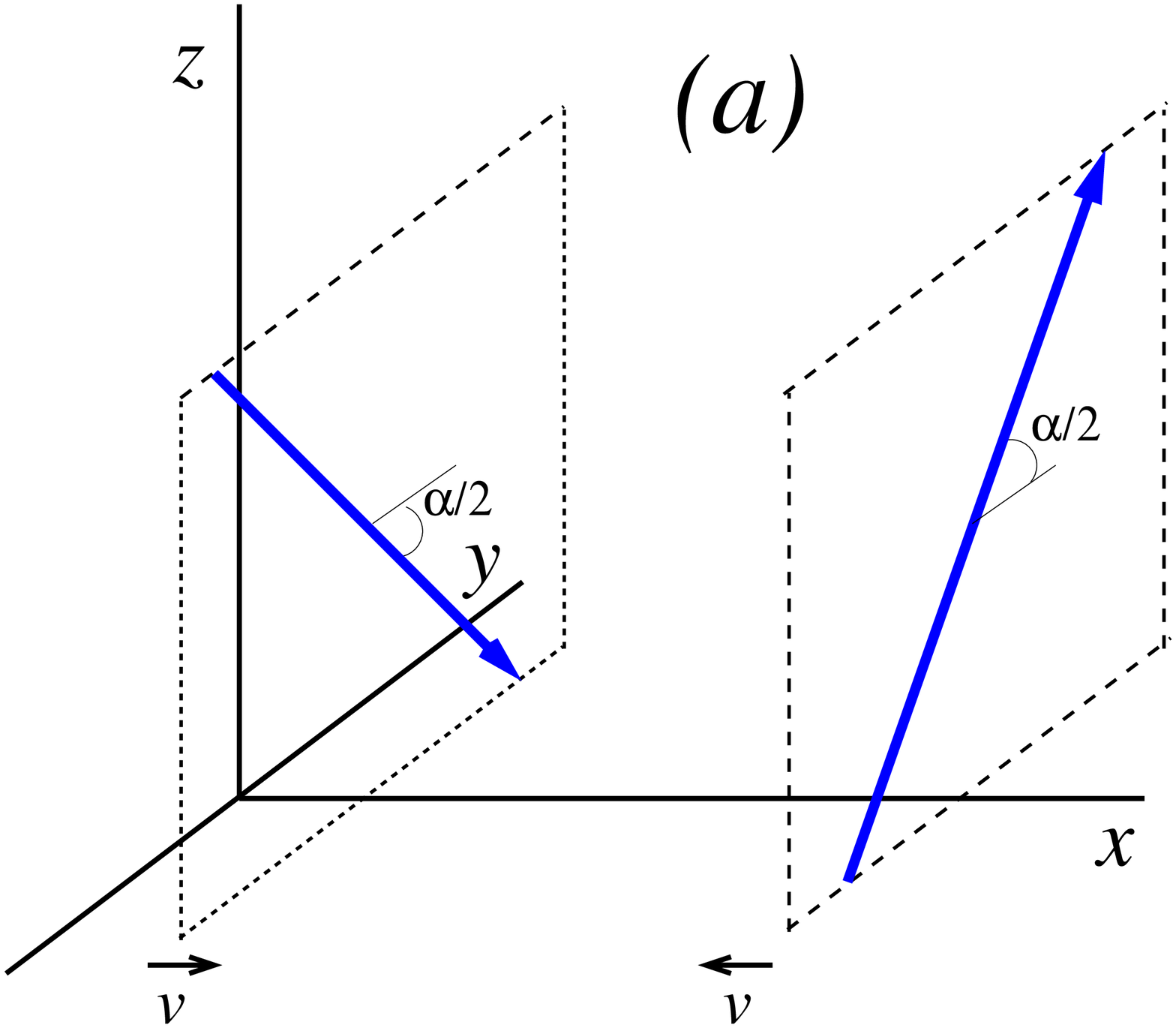}
  \includegraphics*[width=5.5cm]{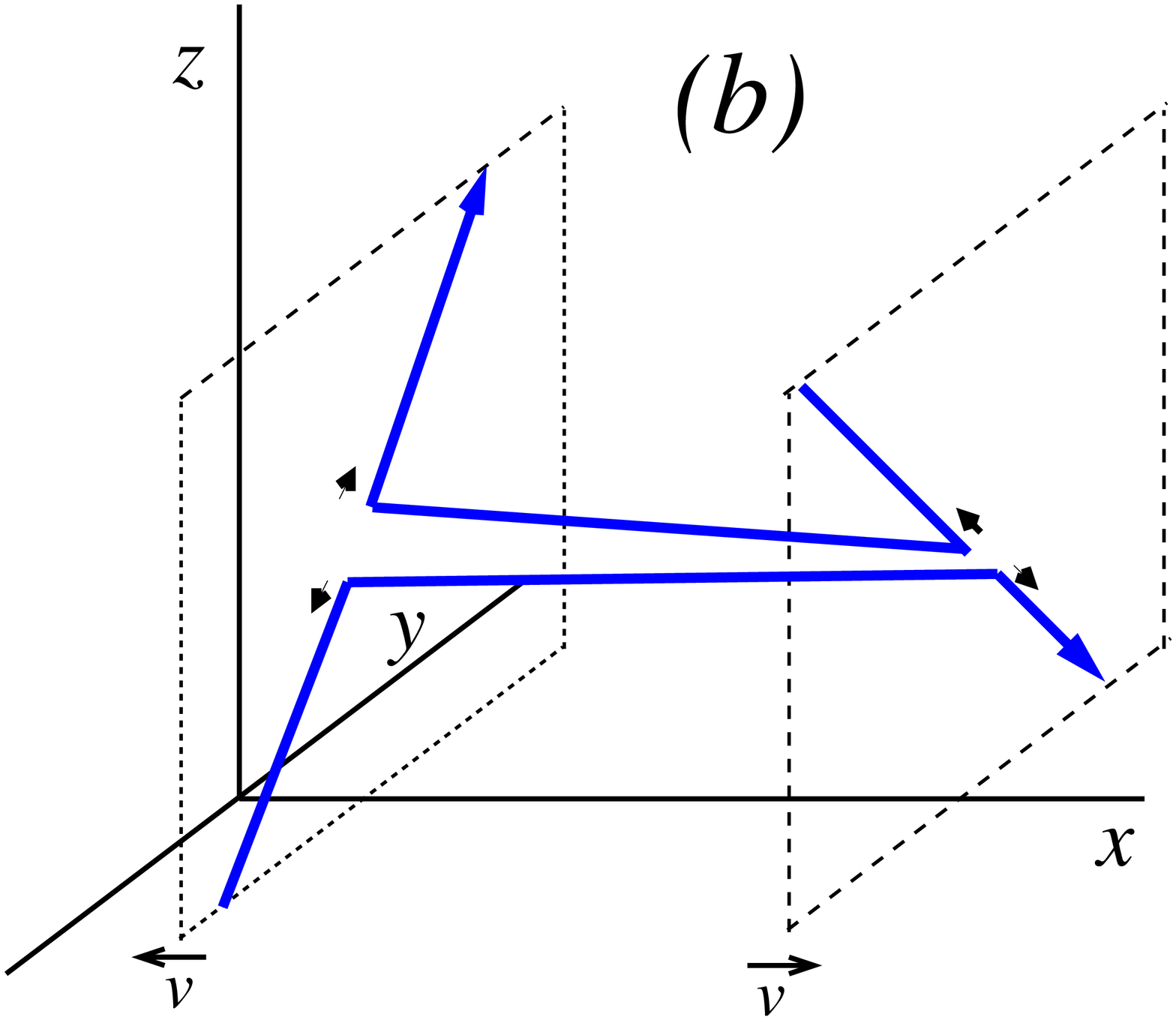}
  }
  \end{center}
  \caption{(a) Initial positions and orientations of the strings
  in the center of mass frame.  The strings lie  in $x =$
  const.  planes and approach each other with speed $v$.
The arrows indicate the orientations of the strings, which form an
  angle $\alpha$.
(b): The configuration after one intercommutation.
If $v \sim c$, the kinks' motion along the strings is
  negligible and the connecting horizontal segments are practically
  antiparallel and immobile, making a second interaction possible.  }
  \label{in conf}
\end{figure}

In this paper we focus on the intercommutation behavior of
ANO strings in the abelian Higgs model --the relativistic
Landau-Ginzburg model-- but our results should apply to
other abelian local (gauged) strings
provided there are no topological obstructions.
Since intercommutation is a local effect 
we work in flat space-time. In units $c = \hbar = 1$,
the lagrangian is 
\begin{equation}
\label{abelHiggslagr}
\mathcal{L} = (\pa_{\mu} +  \imag e A_{\mu}) \phi (\pa^{\mu} - \imag e A^{\mu})\phi^{\dagger} - \frac{1}{4}F^{\mu \nu}F_{\mu 
\nu} - \frac{\lambda}{4} (|\phi|^2 - \eta ^2)^2.
\end{equation}
After a rescaling of units
this model can be
characterized by a single parameter, the ratio of the
Higgs mass to the gauge boson mass $\beta :=
({m_{\phi}}/{m_{A}})^2 = {\lambda}/{2 e^2}$. Following
\cite{Matz88,Shellard87}, we  place a superposition of
two (boosted) 
ANO strings on a three dimensional lattice
and evolve this configuration in the Hamiltonian formalism
using a leapfrog algorithm. 
The initial configuration is characterized by only two parameters (see
fig.  \ref{in conf}(a)): the center of mass speed $v$ of the strings
when they are far apart and the angle $\alpha$ between them.

\vspace{0.5cm}

The main conclusion from previous simulations (e.g. \cite{Matz88}) of
the interaction of ANO
strings with unit winding is that
intercommutation takes place in all cases. 
This can be understood by looking at the field configuration in
certain two dimensional slices through the point, say ($x_0,y_0,z_0$),
in which the strings come to intersect \cite{VilShell94}. In the $z =
z_0$ plane (see figure \ref{in conf}), the string interaction looks
like the collision and annihilation of a vortex and an antivortex,
whereas in the $y = y_0$ plane it looks like vortex - vortex
scattering at
$90^o$ \cite{MyersRebbiStrilka92}.
Hence, large-scale network simulations of ANO strings use the
Nambu-Goto approximation and assume that strings {\it always} exchange
ends when they collide.
Here we argue that this picture has to be modified  for sufficiently high
collision speeds.  We find evidence of a threshold speed $v_t$ above
which strings exchange ends {\it twice} and thus effectively pass
through each other.  Note that this is a different effect than the
threshold speed suggested elsewhere for the first intercommutation
\cite{Cope86,Shellard87,Han05}, of which we find no evidence.
We will now first describe some of the more technical aspects of our
simulations and then explain our results in some detail.

\vspace{0.5cm}

{\bf Simulations and Results}

\vspace{0.5cm}

For each simulation, the lattice spacing $a$ is determined as
follows. The vortex configuration
\cite{Abr57,NielOl73} is $\phi = \eta \, X(r) e^{\imag
\theta}$ in cylindrical polar coordinates, with $X(0) = 0$ and
$X(\infty) = 1$, such that $(X'(0))^{-1}$ gives a characteristic
scale for the (Higgs) core.
For $\beta = 1/8,1$, we always take the lattice spacing to be $a \approx (5 \gamma(v) X'(0))^{-1}$,
where $\gamma(v) = 1/\sqrt{1-v^2}$, while for $\beta = 8,32$, we sometimes
have to settle for slightly less resolution (but $a$ is never larger than
twice the size given above).
The typical time step size is
$\Delta t \approx a/2$, so the Courant condition (here $\Delta t \leq
a/\sqrt{3}$) holds, and most simulations are performed on a $400^4$ grid.
The initial string separation is taken to be
$\Delta x \approx 2 R/\gamma(v)$, where $R$ is the radius of a
stationary string/vortex outside of which both the Higgs and the gauge
fields are within $5 \%$ of their vacuum values (this means that $v$
is not exactly what it would be at infinite separation, but the
difference is negligible).
We use the same boundary conditions as
\cite{Matz88}: after each round the fields inside the box are updated
using the equations of motion, and the fields on the boundaries are
calculated
assuming the strings move unperturbed and at constant speeds at the
boundaries.

\vspace{0.5cm}

\begin{figure}[t]
  \begin{center}{
  \includegraphics[width=6cm]{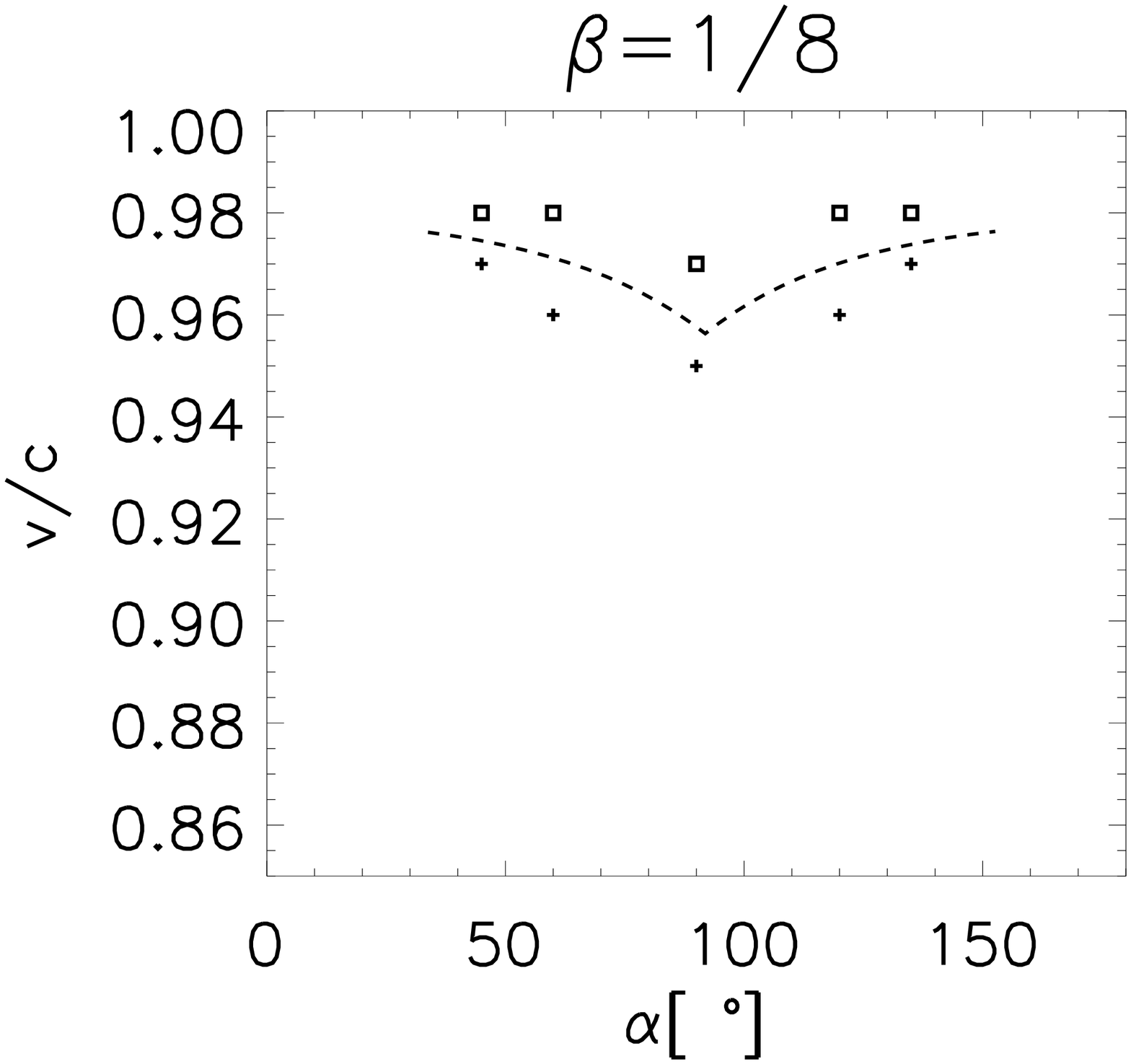}
  \includegraphics[width=6cm]{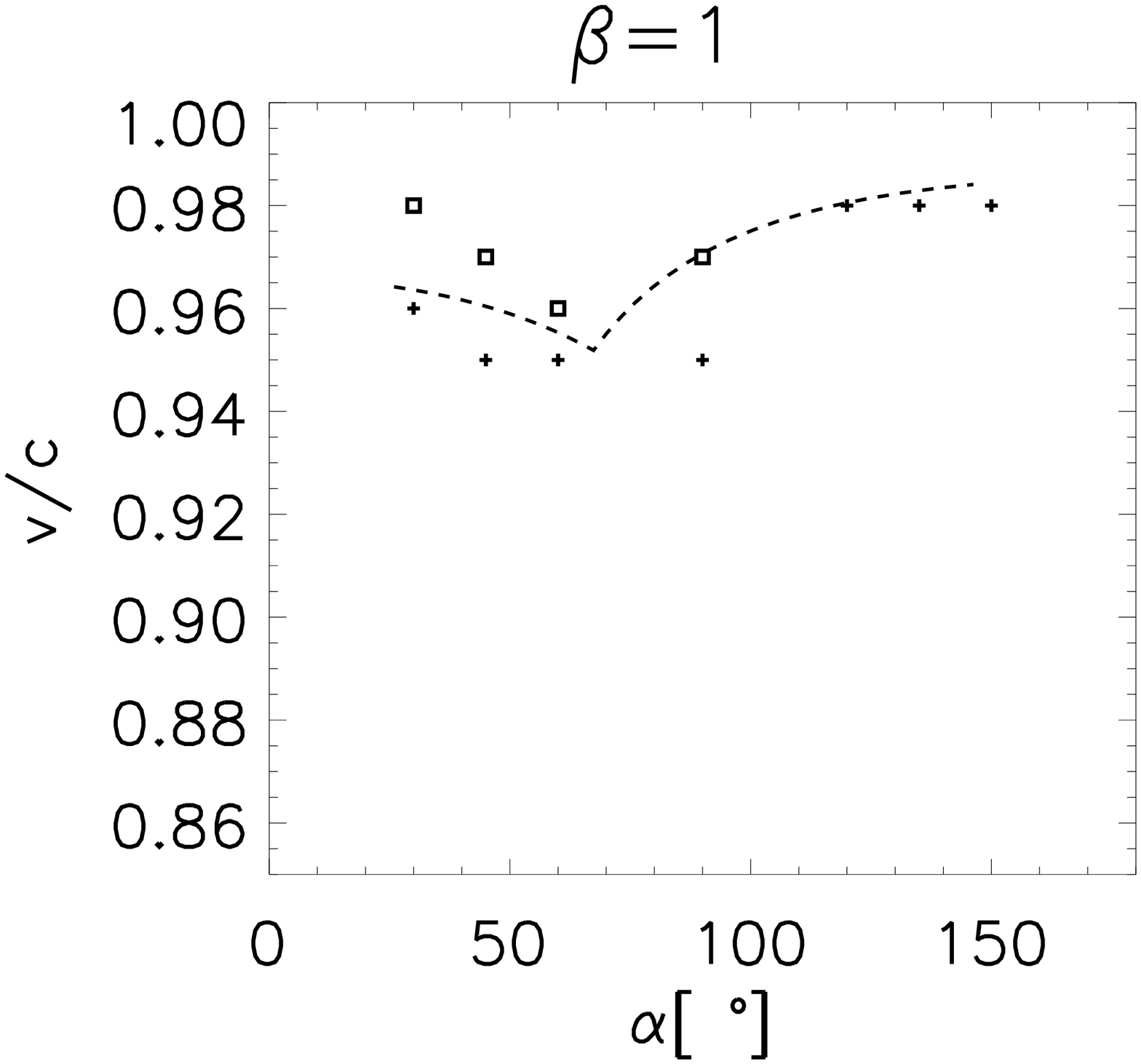}
  }
  \end{center}
  \caption{Threshold speed as a function of incidence angle
for $\beta = 1/8, 1$.
For each $\alpha$, a dot gives the {\it highest} approach speed
$v$ (in our simulations) for which strings only exchange ends once,
and a square gives the {\it lowest} speed for which strings reconnect
twice (so the threshold speed $v_t$ is in between).
Dashed lines are based on a simple theoretical model (see text;
the plots shown have $\delta_t = 156^o, 150^o$ and $w_t =
0.18,0.17$ for $\beta = 1/8, 1$ resp.).  }
  \label{Results0}
\end{figure}

\begin{figure}[t]
  \begin{center}{
  \includegraphics[width=6cm]{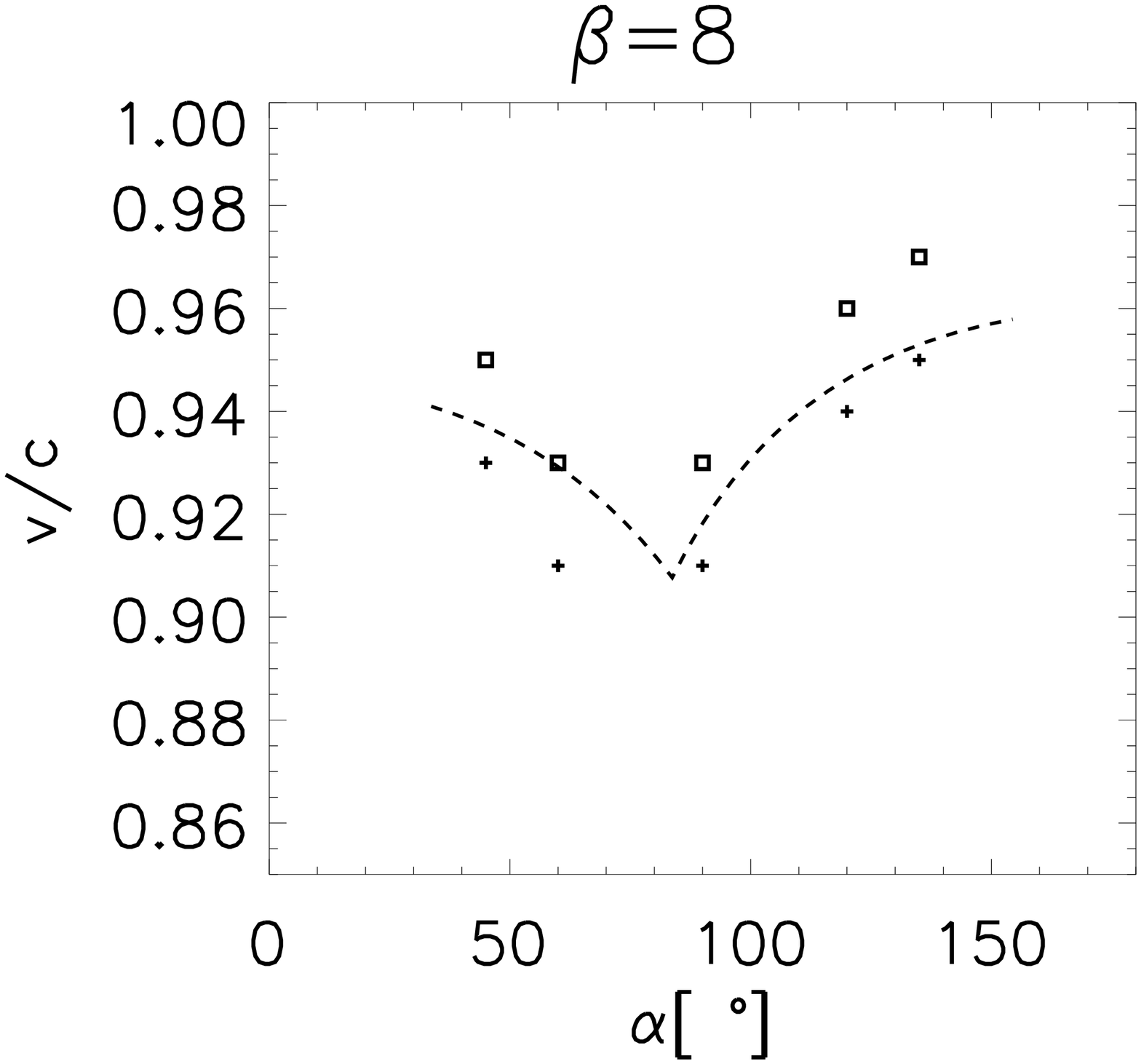}
  \includegraphics[width=6cm]{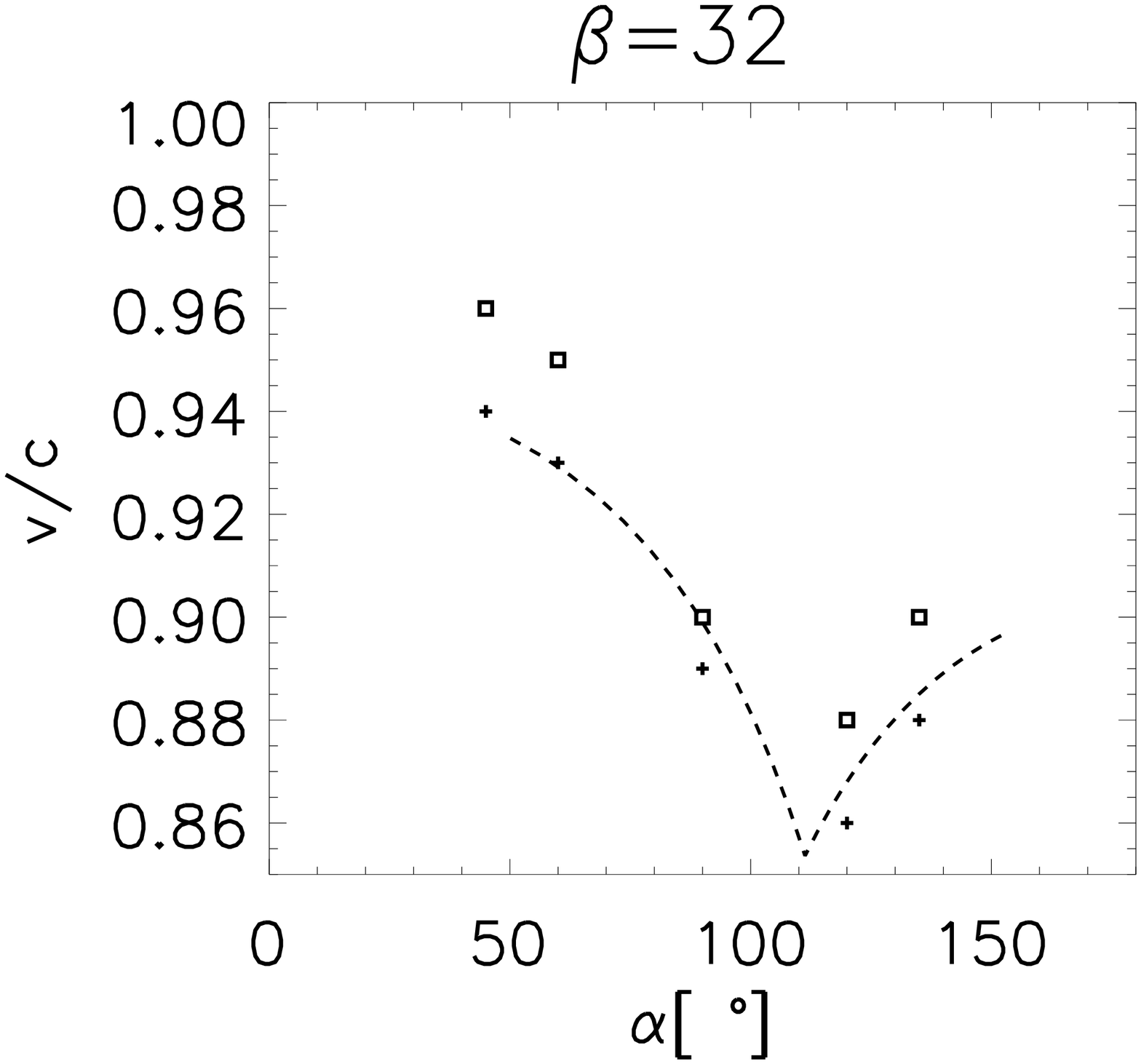}
  }
  \end{center}
  \caption{ Threshold speeds for type II strings. For very large
  $\beta$ and $\alpha$, a different mechanism governs the second
  intercommutation (interaction with an emerging string loop) and the
  threshold speeds are considerably smaller than
  for lower $\beta$.
  However, the results still agree well with our model. We simply
  need a higher critical speed $w_t$ for $\beta = 32$.
  Fits use $\delta_t = 142^o, 142^o$ and $w_t =
  0.28,0.43$ for $\beta = 8, 32$ respectively.
  }
  \label{Results1}
\end{figure}

\begin{figure}
  \begin{center}{
  \includegraphics[width=5cm]{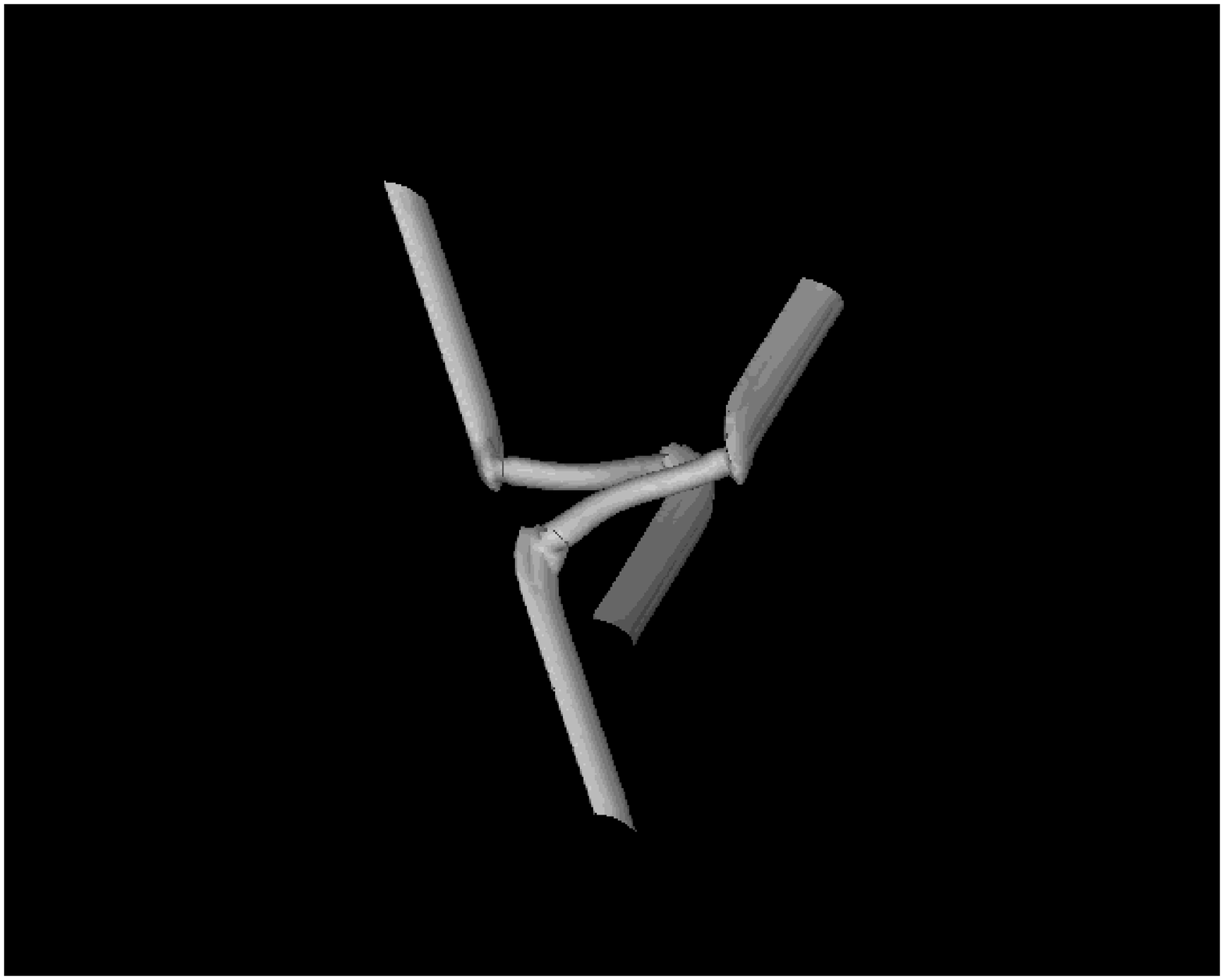}
  \includegraphics[width=5cm]{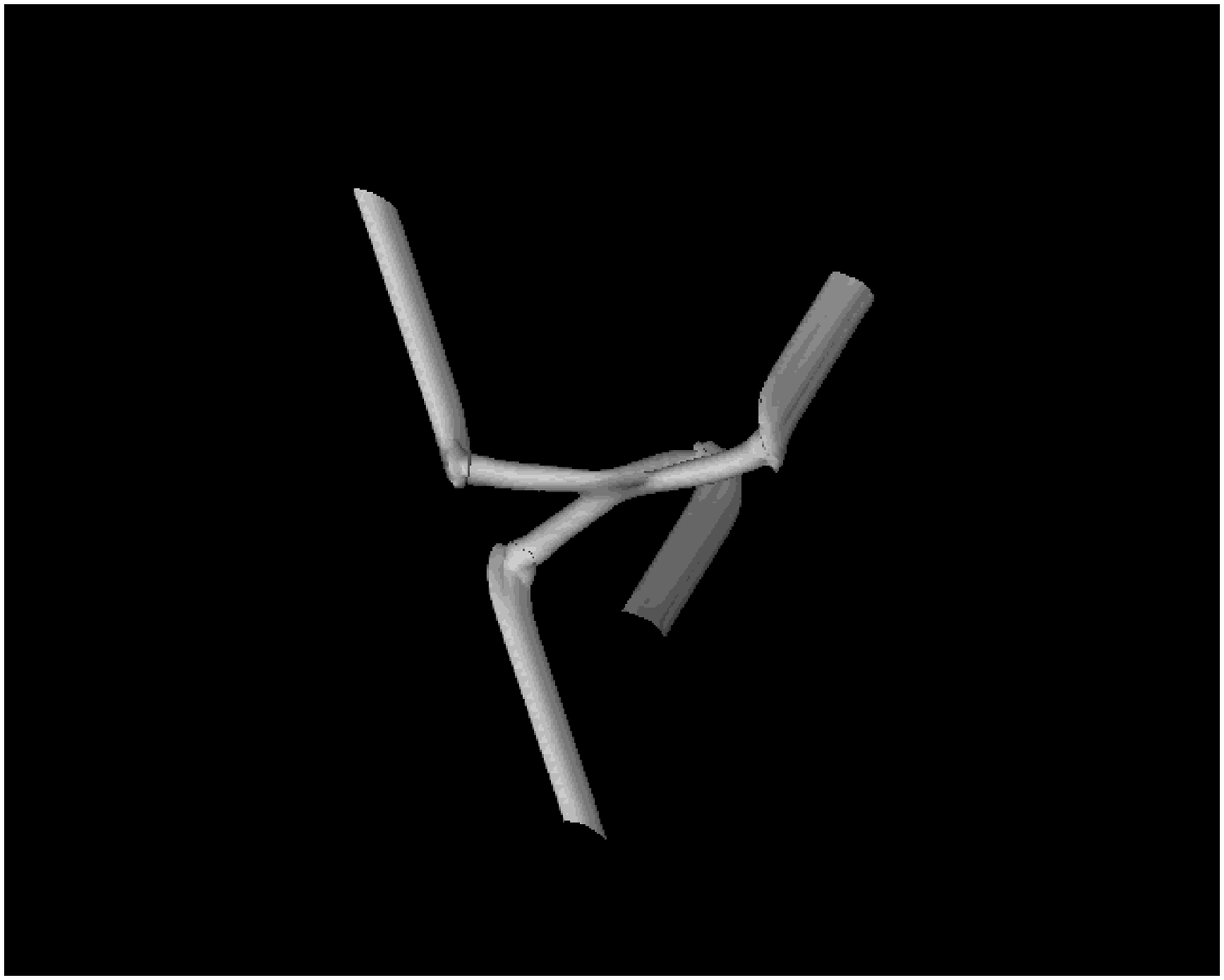}
  \includegraphics[width=5cm]{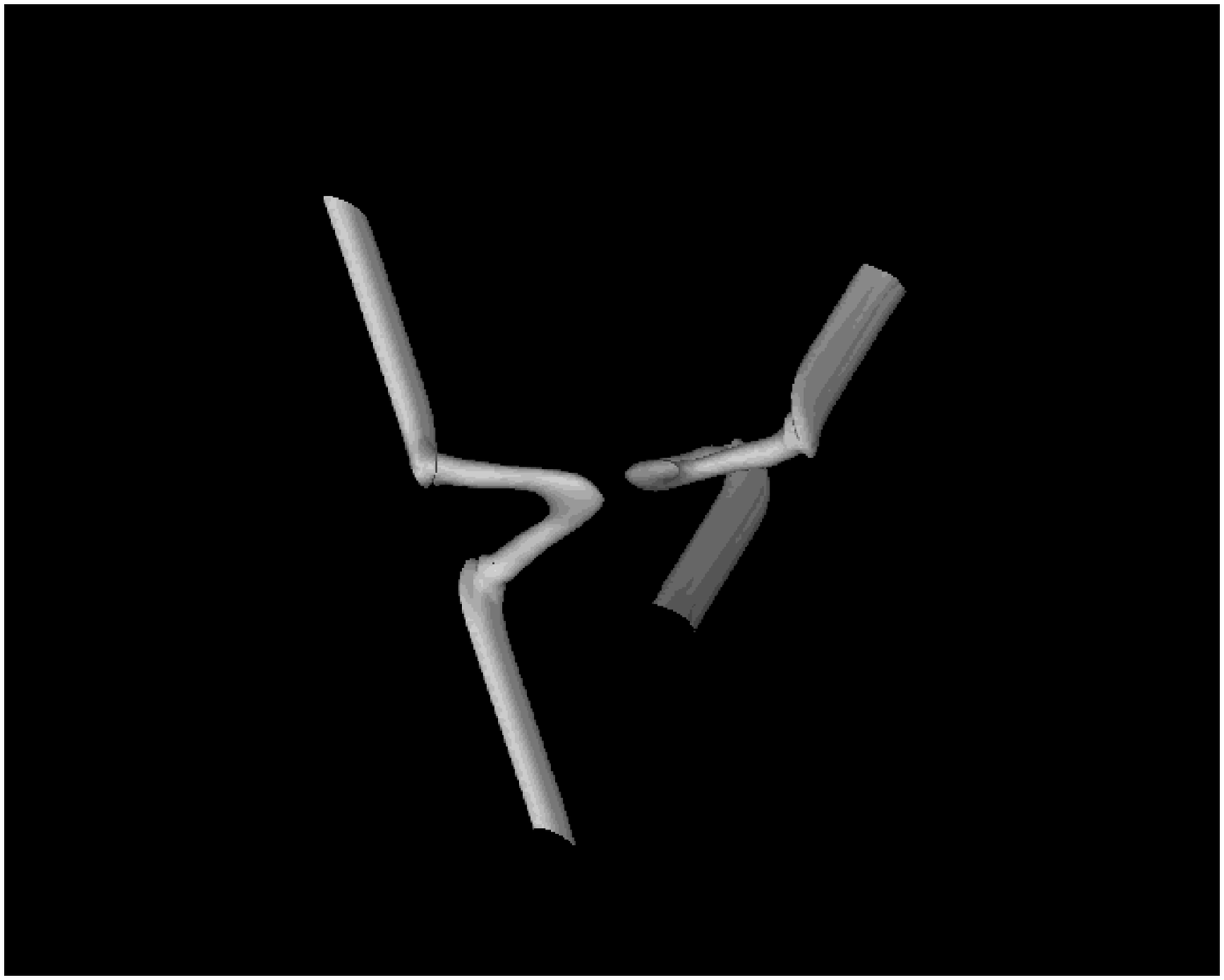}
  }
  \end{center}
  \caption{Double intercommutation of strings with $\beta = 1$, 
$\alpha = 60^o$ and $v =
  0.96$.
The strings effectively pass through each other, with some distorsion.
We use a box size $35 \times 35 \times 36.4$.
Within the shaded surfaces, the energy density is
above $\sim 10 \%$ of the peak energy density of a
{\it static} 
ANO string/vortex.  Fast moving segments
appear to be thicker. 
The strings first intercommute ($T \sim 1.0$, not shown) and separate
(top image, $T = 7.2$). Next, they attract, come to intersect again
(center, $T = 9.0$), intercommute a second time and move away from
each other (bottom, $T = 10.8$). We use length and time units which take $e = \eta = 1$.}
  \label{Results2}
\end{figure}

The results of our simulations are presented in figures \ref{Results0}
and \ref{Results1}. We investigate values of $\beta = 1/8,1,8,32$ and
$\alpha = 30^o,45^o,60^o,90^o,120^o,135^o,150^o$ and center of mass speeds up to
$v = 0.98$ and find threshold speeds in most cases. We suspect that if
we could probe even higher initial speeds, we would find a threshold
speed in all cases.  For $v < v_t$, the strings intercommute and then
move away from each other without interacting again. 
However, for $v > v_t$ the interaction of the strings after
intercommutation is such that they exchange ends a second time!
The exact mechanism through
which the strings reconnect the second time depends on the value of
$\beta$ and $\alpha$. In most cases (always for $\beta = 1/8$ and
$\beta = 1$), the process unfolds roughly as in figure \ref{Results2}:
the strings attract after the first
intercommutation, come to intersect again in the center of the box and
then intercommute a second time. Afterwards, the strings move on as if
they had simply passed through each other, except that the parts of
the strings that have been involved in the interaction lag behind the
rest of the strings a little.
For $\beta>>1$ (deep type II, e.g. $\beta = 32$) and large initial
angle $\alpha = 135^o$, the second exchange of ends proceeds
differently. In this case, a string loop is formed (the three
dimensional manifestation of vortex - antivortex reemergence, see also
\cite{MyersRebbiStrilka92,Matz88}) after the first
intercommutation. The loop starts at the intersection point and grows
in the $y = y_0$ plane (see fig. \ref{in conf}) to eventually catch up
with the two original strings. When this happens, the strings
reconnect again through the loop and move on as if they have passed
through each other. For certain values of $v < v_t$, we also find loop
formation but the loop does not grow large enough to catch up with the
strings.  Note that our results agree with the conventional picture
of string interactions in the sense that, initially, the strings
{\it always} intercommute.

\vspace{0.5cm}

{\bf Discussion and Outlook}

\vspace{0.5cm}

The existence of a threshold speed and its angle dependence is easily
explained in the Nambu-Goto approximation.  Immediately after the
first intercommutation each string has two kinks separating the parts
unaffected by the 
interaction from the straight horizontal segments created in
between (see fig.\ref{in conf}(b)).
These kinks have to move at the speed of light, and their vertical
motion along the strings pulls the horizontal segments apart. But for
high collision speeds $v\sim c$, the kinks' {\it horizontal} velocity
is almost luminal and so the vertical component negligible; moreover the
horizontal segments are almost antiparallel (in the unphysical limit
$v = c$, the kinks do not move up at all and the segments
are exactly antiparallel and lying on top of each other; so we would
expect the second reconnection with probability one).

\vspace{0.5cm}

More precisely, the angle between the horizontal segments is
$\cos(\delta/2) = \frac{\cos(\alpha/2)/(v\gamma(v))}{\sqrt{1 +
(\cos(\alpha/2)/(v\gamma(v)))^2}}$ (antiparallel for $\delta = \pi$) and
they move apart with velocity $w = {\sin(\alpha/2)}/{\gamma(v)}$, the
vertical velocity of the kinks.  If we assume that strings
intercommute a second time only for  $\delta$ above a critical
angle $\delta_t$
and $w$ below  a critical speed $w_t$,
we can get a surprisingly good fit to the data. First, for $\delta_t
\sim 150$ and $w_t \sim 0.2$ ($w_t \approx 0.43$ significantly higher
for $\beta = 32$ where the second exchange of ends occurs through a loop
for large $\alpha$) we get a threshold speed of about the right
magnitude that does not depend strongly on $\alpha$. Second, when we
do look at the $\alpha$ dependence, the model nicely explains the
minimum of $v_t$ somewhere around $\alpha = 90^o$.
The heuristic picture is that if the attractive interaction energy of the
horizontal segments exceeds their kinetic energy they will come
together and annihilate again. What the Nambu-Goto approximation
misses is the interaction energy of the bridging segments.

\vspace{0.5cm}

In conclusion, we find that ANO
strings effectively pass through each other at high speeds of
approach. The result is consistent with a simple kinematic argument
so we expect it to apply to any other local (gauged) string as long as there
is no topological obstruction to intercommutation. In particular we
have preliminary evidence that multiple-winding type I strings
also pass through each other at these
high collision speeds.  An interesting open question is whether
strings carrying zero modes or bound states will also pass through
each other at very high collision speeds. In the particular case of
semilocal strings and skyrmions in the Bogomolnyi limit it has
recently been found \cite{Lagetal06} that at the location
of the first intercommutation the strings revert to Nielsen-Olesen
strings, so we expect the result to hold there as well.

\vspace{0.5cm}

It is difficult at this point to make a reliable estimate of whether
double intercommutation has a significant effect on the network's
evolution and scaling properties;
this  may have to be
determined in large numerical simulations. 
Obviously $v_t$ is found to be high, 
so the probability of two very fast moving segments colliding is
extremely low. On the other hand, non-intercommutation affects the
highest energy components of the network, and moreover these fast
string segments are expected to have higher collision rates
simply because they cross larger distances, so the question
is definitely worth a second look.

\vspace{0.5cm}

We thank Jon Urrestilla for advice on computational
issues and for very helpful discussions, and also Kepa Sousa, Petja Salmi and Tanmay Vachaspati. 
Our work
was partly supported by the Netherlands Organization for Scientific
Research (N.W.O) under the VICI programme. We acknowledge the generous
allocation of computational resources by the Dutch National Computer
Facility Foundation (NCF).

\bibliography{strings}

\end{document}